\documentclass{emulateapj}

\usepackage{graphicx}
\usepackage{epsfig}
\usepackage{times}
\usepackage{natbib}
\usepackage{url}
\usepackage{color}
\usepackage{amssymb,amsmath}

\newcommand{\an}{{Astron. Nachr.}}
\newcommand{\cjaa}{{Chin. J. Astron. Astrophys.}}

\shorttitle{Sympathetic Eruptions}
\shortauthors{T\"or\"ok et al.}

\begin{document}

\title{A model for magnetically coupled sympathetic eruptions}

\author{
T. T\"or\"ok$^{1}$,
O. Panasenco$^{2}$,
V.~S. Titov$^{1}$,
Z. Miki\'c$^{1}$,  
K.~K. Reeves$^{3}$ , 
M. Velli$^{4}$,
J.~A. Linker$^{1}$
AND
G. De Toma$^{5}$}
\affil{$1$ Predictive Science, Inc., 9990 Mesa Rim Rd., Ste. 170, San Diego, CA 92121, USA}
\affil{$2$ Helio Research, La Crescenta, CA 91214, USA}
\affil{$3$ Harvard-Smithsonian Center for Astrophysics, 60 Garden Street, Cambridge, MA 02138, USA}
\affil{$4$ Jet Propulsion Laboratory, California Institute of Technology, Pasadena, CA 91109, USA}
\affil{$5$ HAO/NCAR, P.O. Box 3000, Boulder, CO 80307-3000, USA}

\begin{abstract}
Sympathetic eruptions on the Sun have been observed for 
several decades,
but the mechanisms by 
which one eruption can trigger another one remain poorly understood. We present a 3D MHD 
simulation that suggests two possible magnetic trigger mechanisms for sympathetic eruptions. 
We consider a configuration that contains two coronal flux ropes located within a pseudo-streamer 
and one rope located next to it. A sequence of eruptions is initiated by triggering the eruption 
of the flux rope next to the streamer. The expansion of the rope leads to two consecutive 
reconnection events, each of which triggers the eruption of a flux rope by removing a sufficient 
amount of overlying flux. The simulation qualitatively reproduces important aspects of the global 
sympathetic event on 2010 August 1 and provides a scenario for so-called twin filament eruptions. 
The suggested mechanisms are applicable also for sympathetic eruptions occurring in other
magnetic configurations.      
\end{abstract}

\keywords{Sun: corona --- Sun: coronal mass ejections (CMEs) --- Sun: flares --- Sun: filaments, prominences
--- Sun: magnetic topology --- Methods: numerical}

\section{Introduction}
\label{sec:int}
Solar eruptions are observed as filament (or prominence) eruptions, flares, and coronal 
mass ejections (CMEs). It is now well established that these three phenomena are 
different observational manifestations of a {\em single eruption}, which is caused by 
the destabilization of a localized volume of the coronal magnetic field. The detailed 
mechanisms that trigger and drive eruptions are still under debate, and a large number 
of theoretical models 
have
been developed \citep[e.g.,][]{forbes10}.

Virtually all existing models consider single eruptions. The Sun, however, also produces 
{\em sympathetic eruptions}, which occur within a relatively short period of time -- either 
in one, typically complex, active region \citep[e.g.,][]{liuc09} or in different source regions, 
which occasionally cover a full hemisphere \citep[so-called ``global eruptions'';][]{zhukov07}.
It has been debated whether the close temporal correlation between sympathetic eruptions 
is purely coincidental, or whether they are causally linked \citep[e.g.,][]{biesecker00}. Both 
statistical investigations \citep[e.g.,][]{moon02,wheatland06} and detailed case studies 
\citep[e.g.,][]{wang01} indicate that physical connections between them exist\footnote{We 
do not distinguish here between sympathetic flares and sympathetic CMEs, since both are part 
of the same eruption process.}.

The exact nature of these connections has yet to be established. They have been attributed, 
for instance, to convective motions or destabilization by large-scale waves \citep[e.g.,][]
{ramsey66,bumba93}. At present, it seems most likely that the mechanisms by which one 
eruption can trigger another one act in the corona and are of 
a
magnetic nature. Perturbations 
traveling along field lines that connect source regions of eruptions \citep[e.g.,][]{jiang08} 
and changes in the background field due to reconnection \citep[e.g.,][]{ding06,zuccarello09} 
have been considered. In an analysis of a global sympathetic event (see Section\,\ref{sec:obs}), 
\cite{schrijver11} found evidence for connections between all involved source regions via 
structural features like separators and quasi-separatrix layers \citep[QSLs;][]{priest92,
demoulin96}; suggesting the importance of the structural properties of the large-scale 
coronal field in the genesis of sympathetic eruptions.

A magnetic configuration that appears to be prone to producing sympathetic eruptions is a 
unipolar streamer or pseudo-streamer \citep[PS; e.g.,][]{hundhausen72,wang07a}. A PS is 
morphologically similar to a helmet streamer, but divides open fields of like polarity and 
contains an even number (typically two) of closed flux lobes below its cusp. PSs are quite 
common in the corona \citep[e.g.,][]{eselevich99,riley11} and occasionally harbor two 
filaments. It seems that if one of these erupts, the other one follows shortly thereafter 
\citep[so-called ``twin filament eruptions'';][]{panasenco10}.

Here we present a numerical simulation that suggests two possible magnetic trigger mechanisms 
for sympathetic eruptions. It was inspired by the global sympathetic event on 2010 August 1, 
which involved a twin filament eruption in a PS.

\begin{figure*}[t]
\centering
\includegraphics[width=1.0\linewidth]{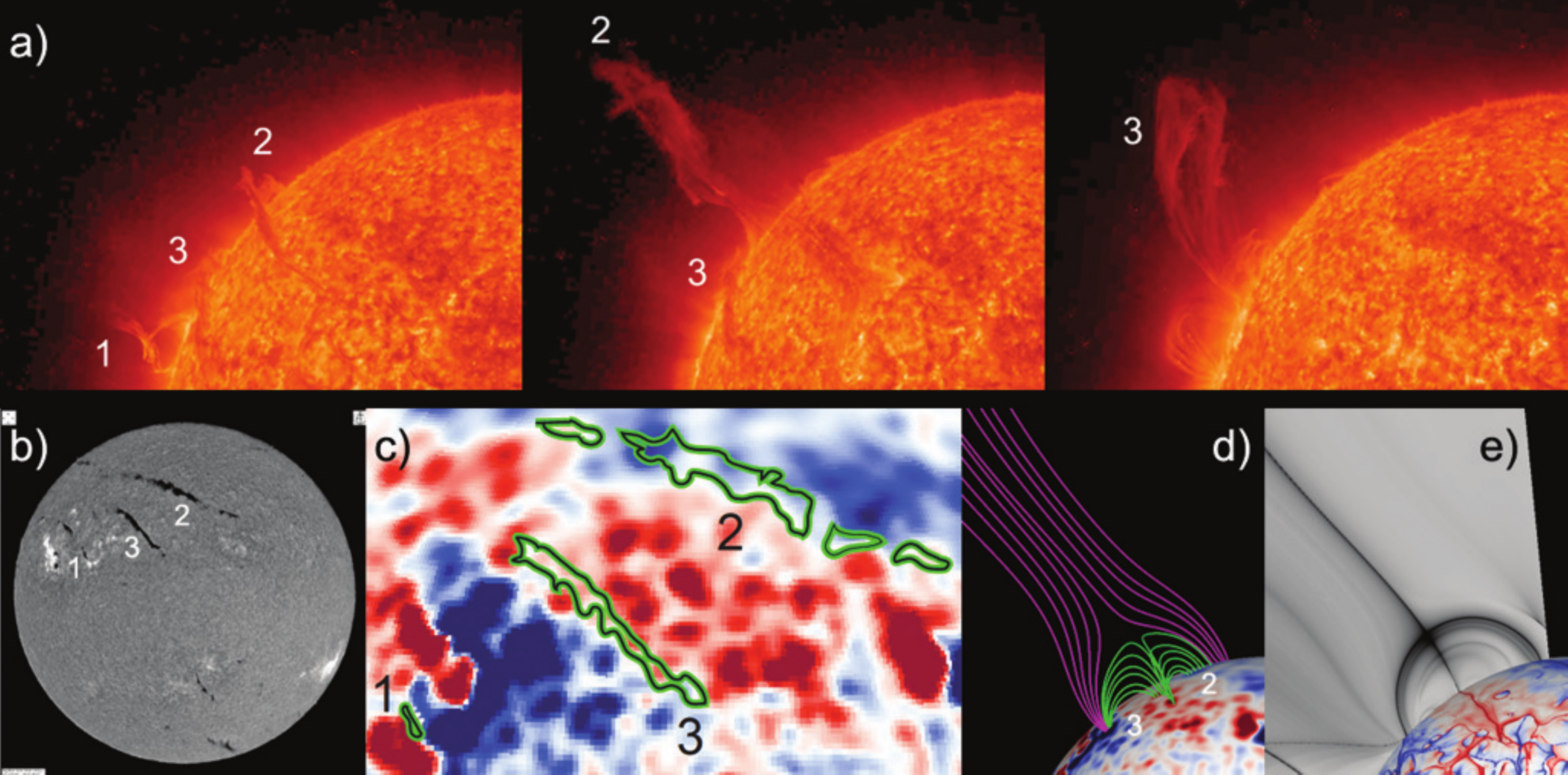}
\caption
{
(a:) STEREO-Ahead/EUVI 
304 \AA\  observations of three subsequent prominence eruptions 
(marked by their order of eruption)
on 2010 August 1, shown at 02:56, 09:16, and 22:06 UT (from left to right). 
(b): Big Bear Observatory H$\alpha$ observation on 2010 July 30, showing the 
corresponding pre-eruptive filaments.
(c): Filament contours (drawn by eye) overlaid on a synoptic 
magnetogram for Carrington rotation 2099, with red (blue) showing positive (negative) 
radial fields.
(d): Magnetic field lines from a corresponding PFSS extrapolation, revealing a 
pseudo-streamer. Green lines outline the lobes in which filaments 2 and 3 were 
located, pink lines show adjacent coronal holes.
(e): Coronal distribution of $Q$ (grayscale) and photospheric distribution of slog $Q$,
where red (blue) outlines positive (negative) magnetic fluxes.
%
%
}
\label{fig:obs}
\end{figure*}

\section{The sympathetic eruptions on 2010 August 1}
\label{sec:obs}
A detailed account of the individual eruptions that occurred in this global 
event can be found in \cite{schrijver11}. Here we focus on a subset of three consecutive filament 
eruptions, all of which evolved into a separate CME. 
%
%
Figures\,\ref{fig:obs}a, b, and c show, respectively, the eruptions as seen by STEREO/EUVI 
\citep{howard08}, the pre-eruptive filaments, and a synoptic magnetogram obtained from 
SoHO/MDI \citep{scherrer95} data.
The large filaments 2 and 3 were located along the inversion 
lines dividing an elongated positive polarity and two bracketing negative polarities; the small 
filament 1 was located at the edge of the southern negative polarity. 
A potential field source surface extrapolation \citep[PFSS; e.g.,][]{schatten69} for Carrington 
rotation 2099 reveals that filaments 2 and 3 were located in the lobes of a PS (Figure\,\ref{fig:obs}d; 
see also \citealt{panasenco10}). 

Figure\,\ref{fig:obs}e shows a cut through the coronal distribution of the squashing factor $Q$ 
\citep{titov02} above filaments 2 and 3. The dark lines of high $Q$ outline structural features 
and exhibit here a shape characteristic for a PS (compare with Figure\,\ref{fig:reco}b below).
The photospheric distribution shows slog $Q$ \citep{titov11}, depicting the footprints 
of (quasi-)separatrix surfaces.
The structural skeleton of a PS consists of two separatrix surfaces, one vertical and one dome-like, 
which are both surrounded by a thin QSL \citep{masson09} and intersect at a separator \citep{titov11}. 
It has been demonstrated that current sheet formation and reconnection occur preferably at such 
separators \citep[e.g.,][]{baum80,lau90}. 

The presence of the PS above filaments 2 and 3 suggests that the CME associated with filament 
eruption 1 may have triggered the subsequent eruptions by destabilizing the PS, presumably
by inducing reconnection at its separator. We now describe an MHD simulation that enabled 
us to test this scenario using an idealized model.

\begin{figure*}[t]
\centering
\includegraphics[width=1.0\linewidth]{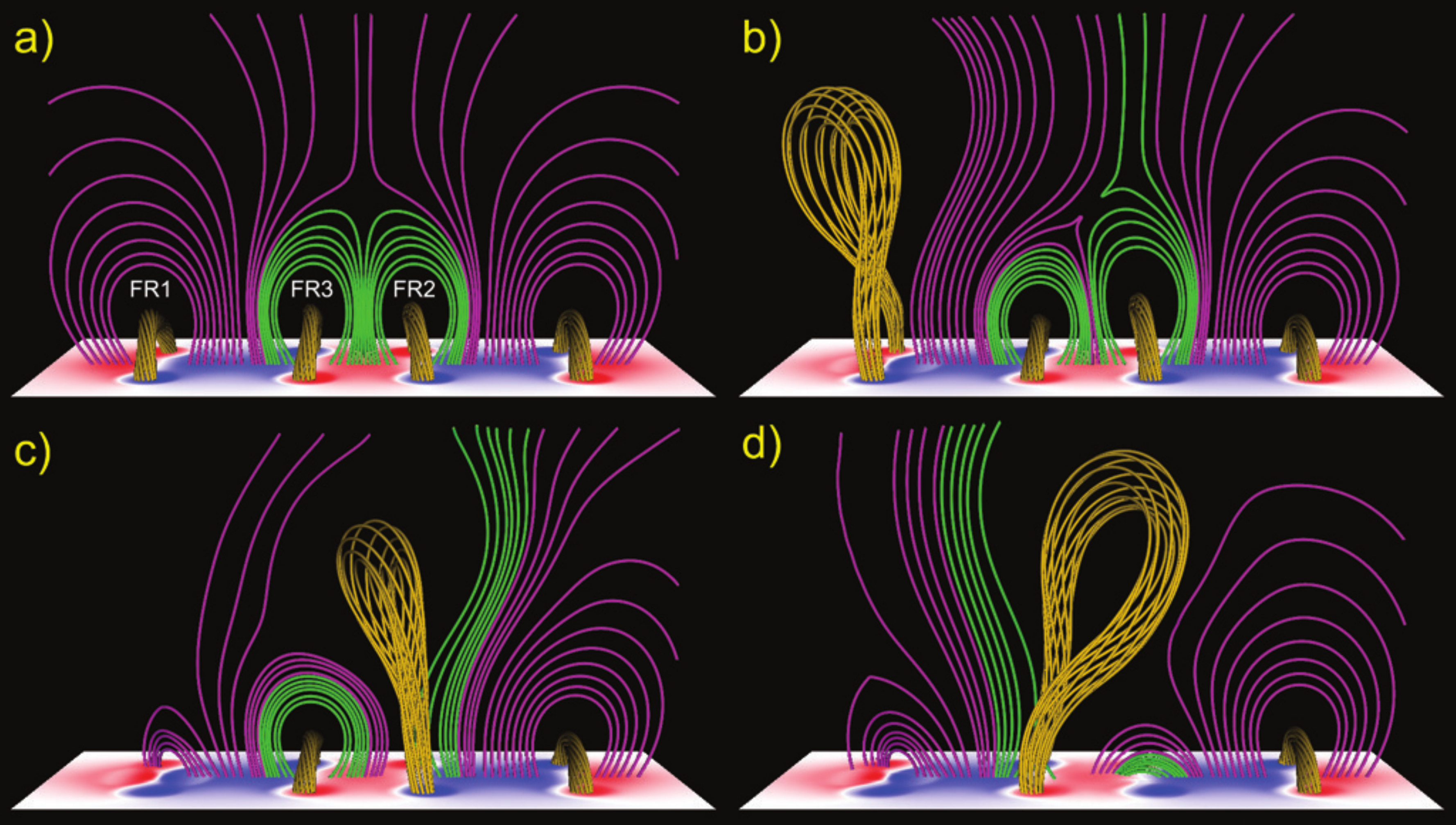}
\caption
{
Snapshots from the simulation, showing magnetic field lines with fixed footpoints and 
the normal component of the magnetic field at the bottom plane, where red (blue) depicts 
positive (negative) fields. Orange lines belong to the flux ropes, green ones to the 
initial pseudo-streamer lobes, and pink ones to initially closed or (semi-)open overlying 
flux. Panel (a) shows the configuration after initial relaxation and (b--d) show the successive 
flux rope eruptions and ambient field evolution at $t=85, 126, \mathrm{and}\,181\,\tau_a$, 
respectively. ``Already erupted'' flux ropes are omitted for clarity.  
(An animation of this figure is available in the online journal.)
}
\label{fig:simu}
\end{figure*}

\begin{figure*}[t]
\centering
\includegraphics[width=1.0\linewidth]{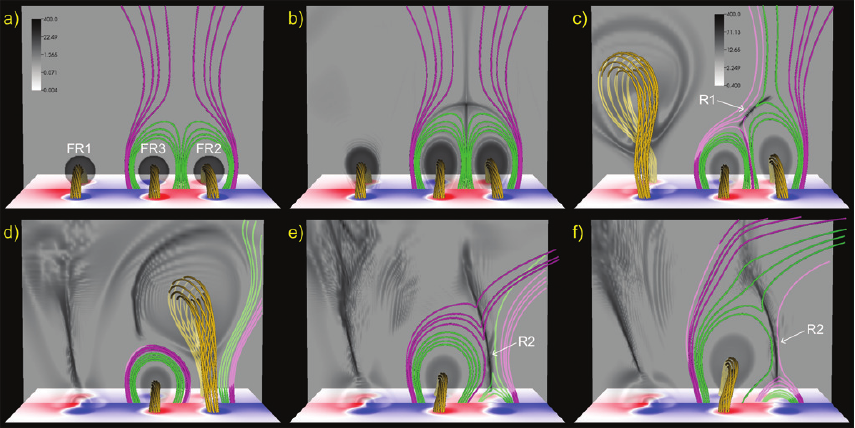}
\caption
{
Illustration of the two reconnection phases that trigger the successive flux 
rope eruptions in the pseudo-streamer. Field lines are colored as in 
Figure\,\ref{fig:simu}. The transparent inverted grayscale in the central plane, 
$\{y=0\}$, shows the logarithmic distribution of $|\mathbf{j}|/|\mathbf{B}|$, 
where $\mathbf{j}$ is the electric current density, outlining flux rope currents 
and thin current layers. Fainter field line segments are located behind the 
transparent layer. Panel (a) shows the initial configuration, (b) the system after 
relaxation, and (c--f) show the dynamic evolution, at $t=85, 126, 142, 
\mathrm{and}\,158\,\tau_a$, respectively. Panels (a,b) and (c--f) use a different 
scaling of $|\mathbf{j}|/|\mathbf{B}|$, respectively. Panel (c) shows reconnection 
R1, which triggers the eruption of FR2, and (e,f) show reconnection R2, which 
triggers the eruption of FR3. Panel (d) shows a state between the two reconnection 
phases. 
(An animation of this figure is available in the online journal.)
}
\label{fig:reco}
\end{figure*}

\section{Numerical simulation}
\label{sec:simu}

The basic simulation setup is as in \cite{torok11}, where two instances of the coronal 
flux rope model by \citet[][hereafter TD]{titov99} were used to simulate the interaction 
of two flux ropes in a PS. Here we add a TD configuration on each side of the PS 
(Figure\,\ref{fig:simu}a). The new configuration on the left is used to model the CME 
associated with filament eruption 1, while the new one on the right is merely used to 
obtain a (line-)symmetric initial configuration, which facilitates the construction of a 
numerical equilibrium. It does not significantly participate in the dynamic evolution 
described below. The flux ropes FR1-3 are intended to model filaments 1-3.
 
We integrate the zero $\beta$ compressible ideal MHD equations, neglecting thermal 
pressure and gravity. The equations are normalized by the initial TD torus axis apex 
height, $R-d$ (see TD), the maximum initial magnetic field strength and Alfv\'en velocity, 
$\boldsymbol{B}_{0\mathrm{max}}$ and $v_{a0\mathrm{max}}$, and derived quantities. 
The Alfv\'en time is $\tau_a=(R-d)/v_{a0\mathrm{max}}$. We use a nonuniform cartesian 
grid of size $[-25,25] \times [-25,25] \times [0,50]$ with resolution $\simeq 0.04$ in 
the flux rope area. The initial density distribution is 
$\rho_0(\boldsymbol{x})=|\boldsymbol{B}_0\,(\boldsymbol{x})|^{3/2}$, such that 
$v_a(\boldsymbol{x})$ decreases slowly with distance from the flux concentrations. 
For further numerical details we refer to \cite{torok03}.

The model parameters are chosen such that all flux ropes are initially stable with respect 
to the helical kink \citep{torok04} and torus instabilities \citep[TI;][]{kliem06}. The ropes 
are placed along the $y$ direction, at $x= \pm 1.5\,\mathrm{and} \pm 5.5$, and have 
identical parameters ($R=2.75$, $a=0.8$, $d=1.75$, $L=0.5$, $q=4.64$; see TD). The 
signs of the sub-photospheric point charges, $\pm q$, are set according to the signs of 
the polarities surrounding filaments 1-3 (Figure\,\ref{fig:obs}c). The half-distance between 
the charges, $L$, is such that the TI can be triggered by a relatively weak perturbation 
\citep[][]{schrijver08}. To obtain a numerically stable initial configuration that contains 
(semi-)open field above the PS lobes, the two charges associated with each flux rope are 
adjusted to $-0.55\,q/0.65\,q$ (for FR1 and FR4) and to $-0.34\,q/0.24\,q$ (for FR2 
and FR3). The twist is chosen left-handed for all ropes to account for the observed 
dextral chirality of filaments 2 and 3 \citep{panasenco10}. 

We first relax the system for $85\,\tau_a$ and reset time to zero. Then we trigger the eruption of FR1 by 
imposing localized converging flows at the bottom plane (as in \citealt{torok11}), which slowly drive the 
polarities surrounding FR1 toward the local inversion line, yielding a quasi-static expansion of the rope's 
ambient field. The flows are imposed for $25\,\tau_a$ (including phases of linear increase [decrease] to 
a maximum velocity of $0.02\,v_{a0\mathrm{max}}$ [to zero], each lasting $5\,\tau_a$). 

Though we solve the ideal MHD equations, extra diffusion is introduced by numerical 
differencing (as in every MHD code that models solar magnetic fields). This numerical
diffusion is localized in regions where the current density is largest, and leads to 
reconnection of magnetic field lines. Although it is much larger than the diffusion 
expected on the Sun, experience has shown that simulations produce solutions with 
physically expected behavior, as long as the numerical diffusion is sufficiently small. We therefore 
expect that our simulation indicates the true evolution of the system, but that the 
reconnection rates might differ from those present on the Sun.

\section{Results}
\label{sec:res}

Figures\,\ref{fig:simu} and \ref{fig:reco} summarize the main dynamics and 
reconnection occurring in the simulation. Figure\,\ref{fig:reco}a shows the 
initial configuration and Figures\,\ref{fig:simu}a and \ref{fig:reco}b show 
the system after relaxation, during which weak current layers form at the PS 
separatrix surfaces, but no noticeable reconnection occurs. Note the 
correspondence between the current layer pattern and the $Q$-distribution 
shown in Figure\,\ref{fig:obs}e.

As the converging flows are applied, FR1 starts to rise slowly, in response 
to the quasi-static expansion of its ambient field. In contrast to other 
simulations, where such flows have been used to create a flux rope from a 
sheared arcade \citep[{e.g.},][]{amari00}, they do not lead here to noticeable 
reconnection. The slow rise lasts until the rope reaches the critical height 
for TI onset at $t \approx 40\,\tau_a$, after which it rapidly accelerates 
upward driven by the instability \citep[][]{torok07,fan07,schrijver08,aulanier10}. 
FR1 attains a maximum velocity of $\approx 0.45\,v_{a0\mathrm{max}}$ at 
$t \approx 90\,\tau_a$ before it slowly decelerates. Figure\,\ref{fig:simu}b 
shows the system in the course of this eruption. The rise of the rope is 
slightly inclined, due to the asymmetry of its ambient field \citep[e.g.,][]
{filippov01}. The rope rotates counterclockwise about its rise direction (as 
seen from above), due to the conversion of its twist into writhe 
\citep[e.g.,][]{green07}.

\begin{figure}[t]
\centering
\includegraphics[width=1.0\linewidth]{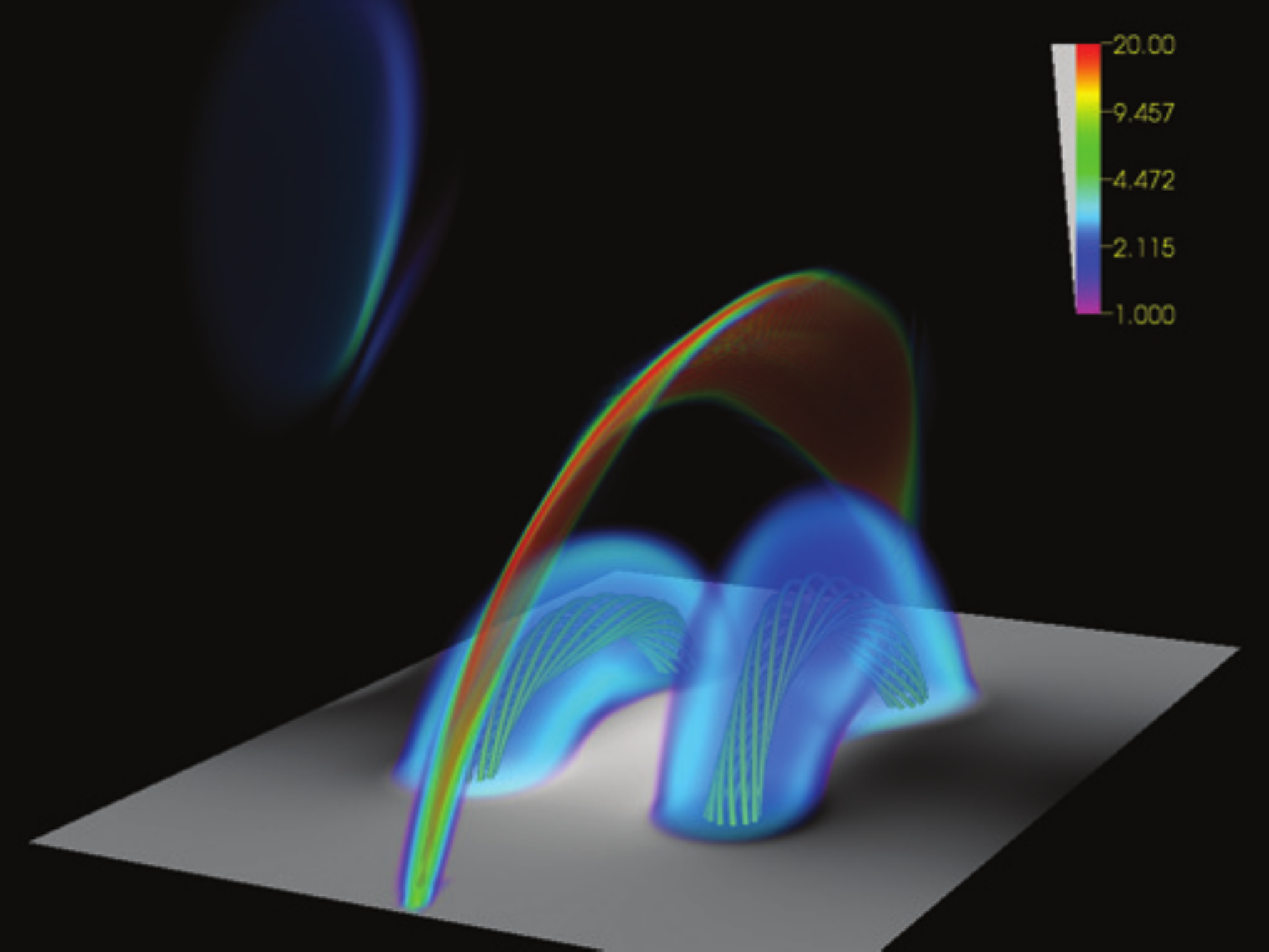}
\caption
{
Volume rendering of $|\mathbf{j}|/|\mathbf{B}|$ in the pseudo-streamer 
area at the same time as in Figures\,\ref{fig:simu}b and \ref{fig:reco}c, 
outlining the tilted arc-shaped current layer that forms 
around
the separator. 
}
\label{fig:pssl}
\end{figure}

The expansion of FR1's ambient field compresses the field between FR1 and the 
PS, particularly at larger heights where it is weak (see online animations). 
As a result, a tilted arc-shaped current layer forms 
around
the PS separator
(Figures\,\ref{fig:reco}c and \ref{fig:pssl}). Further compression by the 
eruption steepens the current densities until reconnection (R1) between the 
open flux to the left of the PS and the closed flux in the right PS lobe sets 
in. The lobe flux then 
starts to open up,
while the open flux 
starts to close
down above the 
left PS lobe (Figures\,\ref{fig:simu}b and \ref{fig:reco}c). This 
successively
decreases (increases) the magnetic tension above FR2 (FR3), so that FR2 rises slowly, 
while FR3 is slowly pushed downward. At $t \approx 95\,\tau_a$ FR2 reaches 
the critical height for TI onset and erupts, attaining a maximum velocity of 
$\approx 0.60\,v_{a0\mathrm{max}}$ at $t \approx 120\,\tau_a$. 
Figure\,\ref{fig:simu}c shows that FR2 also rises non-radially, but rotates 
less than FR1. The apparently smaller rotation of FR2 is due to the faster 
decay of its overlying field with height, which leads to a distribution of 
the total rotation over a larger height range than for FR1 \citep{torok10}. 
By the time shown, FR1 has fully erupted, an elongated vertical current layer 
has formed in its wake (Figure\,\ref{fig:reco}d), and reconnection therein 
has produced cusp-shaped field lines below it. As FR2 erupts, it rapidly pushes 
the arc-shaped current layer to large heights (Figure\,\ref{fig:reco}d). While 
R1 still commences for some time, it does not play anymore a significant role 
for the following evolution. 

A vertical current layer also forms below FR2. The subsequent reconnection (R2) 
initially involves the very same flux systems that took part in R1. The flux 
previously closed down by R1 opens up again, and the flux previously opened up 
by R1 -- and by the expansion of FR2 -- closes down to form cusp-shaped field 
lines below the current layer (Figure\,\ref{fig:reco}e). After these fluxes are 
exhausted, R2 continues, now involving the left PS lobe and the open flux to 
the right of the PS. While the former opens up, the latter closes down as part 
of the growing cusp (Figure\,\ref{fig:reco}f). Thus, R2 continuously removes 
closed flux above FR3. As before, this progressive weakening of magnetic tension 
leads to a slow rise of the rope, followed by its eruption (Figures\,\ref{fig:reco}f 
and \ref{fig:simu}d). The rapid acceleration of FR3 by the TI starts at 
$t \approx 138\,\tau_a$, yielding a maximum velocity of $\approx 0.35\,v_{a0\mathrm{max}}$ 
at $t \approx 175\,\tau_a$. The rope shows a significant rotation and an inclined 
rise which is now mainly directed toward the positive $x$ direction.

\section{Discussion}
\label{sec:dis}
The eruptions of FR2 and FR3 are initiated by the removal of a sufficient
amount of stabilizing flux above the flux ropes via reconnection. R1 is 
similar to quadrupolar ``breakthrough'' or ``breakout'' reconnection 
\citep{syrovatskii82,antiochos99}. Here it is driven by a nearby CME rather 
than by an expanding arcade and, in contrast to the breakout model, a flux 
rope is present prior to eruption. R2, on the other hand, corresponds 
to standard flare reconnection in the wake of a CME. Here it removes flux 
from the adjacent PS lobe, thereby triggering the eruption of FR3. A similar 
mechanism for the initiation of a second eruption in a PS was suggested by 
\cite{cheng05}, who, however, attributed it to reconnection inflows rather 
than to flux removal. We emphasize that R1 and R2 merely {\em trigger} the 
eruptions, which are {\em driven} by the TI and supported by the associated 
flare reconnection \citep[e.g.,][]{vrsnak08}. Thus, in the system studied here, 
both PS eruptions require the presence of a pre-eruptive flux rope. 
We further note that the reconnections do not have to commence for the
whole time period until the TI sets in. It is sufficient if they remove enough 
flux for the subsequently slowly rising flux ropes to reach the critical height 
for TI onset.

R1 is driven by a perturbation of limited duration -- the lateral expansion 
of a nearby CME -- and is slow since it involves only weak fields, around a 
separator at a significant height in the corona. Therefore, its success 
in triggering an eruption depends on parameters like the distance of the CME 
from the PS and the amount of pre-eruptive flux within the PS lobes. Indeed, 
if we sufficiently increase these parameters in the simulation we find that 
R1 still commences, but does not last long enough to trigger an eruption. In 
contrast, R2 is driven by the rise of FR2 and involves strong fields. It is 
therefore faster and more efficient, which supports the finding by 
\cite{panasenco10} that an eruption in one lobe of a PS is often followed by 
an eruption in the neighboring lobe.

Figure \,\ref{fig:simu} shows that the simulation correctly reproduces the 
order of the eruptions shown in Figure\,\ref{fig:obs}a and yields a good match 
of their inclinations and rotations. Assuming that the first eruption indeed 
triggered the subsequent ones, it is surprising that the filament located 
further away from it went off first. While filament 2 may simply have been 
closer to its stability limit than filament 3 (as indicated by its larger height; 
see Figure\,\ref{fig:obs}a), the simulation provides an alternative explanation: 
the perturbation of the separator yields an orientation of the current layer 
that leads to a removal of closed flux only in the right PS lobe 
(Figure\,\ref{fig:reco}c), thus enforcing the eruption of FR2. Hence, although 
we could not find observational signatures of R1 (presumably because the 
involved fields were too weak), the observed eruption sequence supports its 
occurrence. The time intervals between the simulated eruptions exhibit a ratio 
different to the observed ones. Matching the observed ratio requires a search 
for the appropriate model parameters and a more realistic modeling of reconnection, 
which are beyond the scope of this work. 

FR2 reaches a velocity about 35 percent larger than FR1, which is in line 
with \cite{liuy07} and \cite{fainshtein10}, who found that CMEs associated 
with PSs are, on average, faster than those associated with helmet streamers. 
\cite{liuy07} suggested that this difference is due to the typically smaller 
amount of closed flux the former have to overcome. Indeed, FR1 has to pass 
through flux that is closed at all heights above it, while FR2 faces much 
less closed flux, a significant fraction of which is, moreover, removed by 
R1. FR3 remains significantly slower than FR2, most likely because it encounters 
more closed flux at eruption onset, and only partially opened flux later on 
(Figures\,\ref{fig:reco}e,f).

\section{Conclusions}
\label{sec:con}

We present an MHD simulation of two successive flux rope eruptions in a PS, and we 
demonstrate how they can be triggered by a preceding nearby eruption. The simulation 
suggests a mechanism for twin filament eruptions and provides a scenario for a subset 
of the sympathetic eruptions on 2010 August 1. More realistic initial configurations and 
a more sophisticated treatment of reconnection are needed for a quantitative comparison 
with observations.      

Our results support the conjecture that the trigger mechanisms of sympathetic eruptions 
can be related to the structural properties of the large-scale coronal field. However, while 
structural features are present in our model configuration, they do no connect the source 
region of the first eruption with the source regions of the subsequent ones. Moreover, the 
mere presence of such features in a source region is not a sufficient criterion for the 
occurrence of a sympathetic event, even if reconnection at structural features is triggered 
by a distant eruption. The conditions in the source region must be such that the resulting 
perturbation forces the region to cross the stability boundary.

The two trigger mechanisms presented here are independent and applicable also to other
magnetic configurations. Triggering a sympathetic eruption by R1 requires the presence of 
a separator (or null point) above closed flux that stabilizes a pre-eruptive flux rope, which 
can be realized, in the simplest case, in a so-called ``fan-spine'' configuration \citep[e.g.,][]
{antiochos98,pariat09,torok09}. Triggering a sympathetic eruption by R2 requires the 
presence of an adjacent closed flux system overlying a flux rope, which can exist, for 
example, in quadrupolar configurations.

\acknowledgments
We thank P. D\'emoulin, B. Kliem, 
and K. Schrijver for stimulating discussions. The contribution 
of T.T, V.S.T., Z.M., and J.A.L. was supported by NASA's HTP, LWS, 
and SR\&T programs, CISM (an NSF Science and Technology Center), 
and a contract from Lockheed-Martin to Predictive Science, Inc.
O.P. was supported by NASA grant NNX09AG27G, G.D.T. by NASA/SHP 
grant NNH09AK02I, and K.K.R. by contract SP02H1701R from 
Lockheed-Martin to SAO. 
M.V's. contribution was carried out at JPL (Caltech) under a contract with NASA.


\begin{thebibliography}{50}
\expandafter\ifx\csname natexlab\endcsname\relax\def\natexlab#1{#1}\fi

\bibitem[{{Amari} {et~al.}(2000){Amari}, {Luciani}, {Mikic}, \&
  {Linker}}]{amari00}
{Amari}, T., {Luciani}, J.~F., {Mikic}, Z., \& {Linker}, J. 2000, \apjl, 529,
  L49

\bibitem[{{Antiochos}(1998)}]{antiochos98}
{Antiochos}, S.~K. 1998, \apjl, 502, L181

\bibitem[{{Antiochos} {et~al.}(1999){Antiochos}, {DeVore}, \&
  {Klimchuk}}]{antiochos99}
{Antiochos}, S.~K., {DeVore}, C.~R., \& {Klimchuk}, J.~A. 1999, \apj, 510, 485

\bibitem[{{Aulanier} {et~al.}(2010){Aulanier}, {T{\"o}r{\"o}k}, {D{\'e}moulin},
  \& {DeLuca}}]{aulanier10}
{Aulanier}, G., {T{\"o}r{\"o}k}, T., {D{\'e}moulin}, P., \& {DeLuca}, E.~E.
  2010, \apj, 708, 314

\bibitem[{{Baum} \& {Bratenahl}(1980)}]{baum80}
{Baum}, P.~J., \& {Bratenahl}, A. 1980, \solphys, 67, 245

\bibitem[{{Biesecker} \& {Thompson}(2000)}]{biesecker00}
{Biesecker}, D.~A., \& {Thompson}, B.~J. 2000, Journal of Atmospheric and
  Solar-Terrestrial Physics, 62, 1449

\bibitem[{{Bumba} \& {Klvana}(1993)}]{bumba93}
{Bumba}, V., \& {Klvana}, M. 1993, \apss, 199, 45

\bibitem[{{Cheng} {et~al.}(2005){Cheng}, {Fang}, {Chen}, \& {Ding}}]{cheng05}
{Cheng}, J., {Fang}, C., {Chen}, P., \& {Ding}, M. 2005, \cjaa, 5, 265

\bibitem[{{D{\'e}moulin} {et~al.}(1996){D{\'e}moulin}, {Henoux}, {Priest}, \&
  {Mandrini}}]{demoulin96}
{D{\'e}moulin}, P., {Henoux}, J.~C., {Priest}, E.~R., \& {Mandrini}, C.~H.
  1996, \aap, 308, 643

\bibitem[{{Ding} {et~al.}(2006){Ding}, {Hu}, \& {Wang}}]{ding06}
{Ding}, J.~Y., {Hu}, Y.~Q., \& {Wang}, J.~X. 2006, \solphys, 235, 223

\bibitem[{{Eselevich} {et~al.}(1999){Eselevich}, {Fainshtein}, \&
  {Rudenko}}]{eselevich99}
{Eselevich}, V.~G., {Fainshtein}, V.~G., \& {Rudenko}, G.~V. 1999, \solphys,
  188, 277

\bibitem[{{Fainshtein} \& {Ivanov}(2010)}]{fainshtein10}
{Fainshtein}, V.~G., \& {Ivanov}, E.~V. 2010, Sun and Geosphere, 5, 28

\bibitem[{{Fan} \& {Gibson}(2007)}]{fan07}
{Fan}, Y., \& {Gibson}, S.~E. 2007, \apj, 668, 1232

\bibitem[{{Filippov} {et~al.}(2001){Filippov}, {Gopalswamy}, \&
  {Lozhechkin}}]{filippov01}
{Filippov}, B.~P., {Gopalswamy}, N., \& {Lozhechkin}, A.~V. 2001, \solphys,
  203, 119

\bibitem[{{Forbes}(2010)}]{forbes10}
{Forbes}, T. 2010, {Models of coronal mass ejections and flares}, ed.
  {Schrijver, C.~J.~\& Siscoe, G.~L.} (Cambridge University Press), 159--191

\bibitem[{{Green} {et~al.}(2007){Green}, {Kliem}, {T{\"o}r{\"o}k}, {van
  Driel-Gesztelyi}, \& {Attrill}}]{green07}
{Green}, L.~M., {Kliem}, B., {T{\"o}r{\"o}k}, T., {van Driel-Gesztelyi}, L., \&
  {Attrill}, G.~D.~R. 2007, \solphys, 246, 365

\bibitem[{{Howard} {et~al.}(2008){Howard}, {Moses}, {Vourlidas}, {Newmark},
  {Socker}, {Plunkett}, {Korendyke}, {Cook}, {Hurley}, {Davila}, {Thompson},
  {St Cyr}, {Mentzell}, {Mehalick}, {Lemen}, {Wuelser}, {Duncan}, {Tarbell},
  {Wolfson}, {Moore}, {Harrison}, {Waltham}, {Lang}, {Davis}, {Eyles},
  {Mapson-Menard}, {Simnett}, {Halain}, {Defise}, {Mazy}, {Rochus}, {Mercier},
  {Ravet}, {Delmotte}, {Auchere}, {Delaboudiniere}, {Bothmer}, {Deutsch},
  {Wang}, {Rich}, {Cooper}, {Stephens}, {Maahs}, {Baugh}, {McMullin}, \&
  {Carter}}]{howard08}
{Howard}, R.~A., {Moses}, J.~D., {Vourlidas}, A., {Newmark}, J.~S., {Socker},
  D.~G., {Plunkett}, S.~P., {Korendyke}, C.~M., {Cook}, J.~W., {Hurley}, A.,
  {Davila}, J.~M., {Thompson}, W.~T., {St Cyr}, O.~C., {Mentzell}, E.,
  {Mehalick}, K., {Lemen}, J.~R., {Wuelser}, J.~P., {Duncan}, D.~W., {Tarbell},
  T.~D., {Wolfson}, C.~J., {Moore}, A., {Harrison}, R.~A., {Waltham}, N.~R.,
  {Lang}, J., {Davis}, C.~J., {Eyles}, C.~J., {Mapson-Menard}, H., {Simnett},
  G.~M., {Halain}, J.~P., {Defise}, J.~M., {Mazy}, E., {Rochus}, P., {Mercier},
  R., {Ravet}, M.~F., {Delmotte}, F., {Auchere}, F., {Delaboudiniere}, J.~P.,
  {Bothmer}, V., {Deutsch}, W., {Wang}, D., {Rich}, N., {Cooper}, S.,
  {Stephens}, V., {Maahs}, G., {Baugh}, R., {McMullin}, D., \& {Carter}, T.
  2008, \ssr, 136, 67

\bibitem[{{Hundhausen}(1972)}]{hundhausen72}
{Hundhausen}, A.~J. 1972, {Coronal Expansion and Solar Wind}, ed. {Hundhausen,
  A.~J.}

\bibitem[{{Jiang} {et~al.}(2008){Jiang}, {Shen}, {Yi}, {Yang}, \&
  {Wang}}]{jiang08}
{Jiang}, Y., {Shen}, Y., {Yi}, B., {Yang}, J., \& {Wang}, J. 2008, \apj, 677,
  699

\bibitem[{{Kliem} \& {T{\"o}r{\"o}k}(2006)}]{kliem06}
{Kliem}, B., \& {T{\"o}r{\"o}k}, T. 2006, \prl, 96, 255002

\bibitem[{{Lau} \& {Finn}(1990)}]{lau90}
{Lau}, Y.-T., \& {Finn}, J.~M. 1990, \apj, 350, 672

\bibitem[{{Liu} {et~al.}(2009){Liu}, {Lee}, {Karlick{\'y}}, {Prasad Choudhary},
  {Deng}, \& {Wang}}]{liuc09}
{Liu}, C., {Lee}, J., {Karlick{\'y}}, M., {Prasad Choudhary}, D., {Deng}, N.,
  \& {Wang}, H. 2009, \apj, 703, 757

\bibitem[{{Liu}(2007)}]{liuy07}
{Liu}, Y. 2007, \apjl, 654, L171

\bibitem[{{Masson} {et~al.}(2009){Masson}, {Pariat}, {Aulanier}, \&
  {Schrijver}}]{masson09}
{Masson}, S., {Pariat}, E., {Aulanier}, G., \& {Schrijver}, C.~J. 2009, \apj,
  700, 559

\bibitem[{{Moon} {et~al.}(2002){Moon}, {Choe}, {Park}, {Wang}, {Gallagher},
  {Chae}, {Yun}, \& {Goode}}]{moon02}
{Moon}, Y., {Choe}, G.~S., {Park}, Y.~D., {Wang}, H., {Gallagher}, P.~T.,
  {Chae}, J., {Yun}, H.~S., \& {Goode}, P.~R. 2002, \apj, 574, 434

\bibitem[{{Panasenco} \& {Velli}(2010)}]{panasenco10}
{Panasenco}, O., \& {Velli}, M.~M. 2010, AGU Fall Meeting Abstracts, A1663

\bibitem[{{Pariat} {et~al.}(2009){Pariat}, {Antiochos}, \& {DeVore}}]{pariat09}
{Pariat}, E., {Antiochos}, S.~K., \& {DeVore}, C.~R. 2009, \apj, 691, 61

\bibitem[{{Priest} \& {Forbes}(1992)}]{priest92}
{Priest}, E.~R., \& {Forbes}, T.~G. 1992, \jgr, 97, 1521

\bibitem[{{Ramsey} \& {Smith}(1966)}]{ramsey66}
{Ramsey}, H.~E., \& {Smith}, S.~F. 1966, \aj, 71, 197

\bibitem[{{Riley} \& {Luhmann}(2011)}]{riley11}
{Riley}, P., \& {Luhmann}, J.~G. 2011, \solphys, {\em under revision}

\bibitem[{{Schatten} {et~al.}(1969){Schatten}, {Wilcox}, \&
  {Ness}}]{schatten69}
{Schatten}, K.~H., {Wilcox}, J.~M., \& {Ness}, N.~F. 1969, \solphys, 6, 442

\bibitem[{{Scherrer} {et~al.}(1995){Scherrer}, {Bogart}, {Bush}, {Hoeksema},
  {Kosovichev}, {Schou}, {Rosenberg}, {Springer}, {Tarbell}, {Title},
  {Wolfson}, {Zayer}, \& {MDI Engineering Team}}]{scherrer95}
{Scherrer}, P.~H., {Bogart}, R.~S., {Bush}, R.~I., {Hoeksema}, J.~T.,
  {Kosovichev}, A.~G., {Schou}, J., {Rosenberg}, W., {Springer}, L., {Tarbell},
  T.~D., {Title}, A., {Wolfson}, C.~J., {Zayer}, I., \& {MDI Engineering Team}.
  1995, \solphys, 162, 129

\bibitem[{{Schrijver} {et~al.}(2008){Schrijver}, {Elmore}, {Kliem},
  {T{\"o}r{\"o}k}, \& {Title}}]{schrijver08}
{Schrijver}, C.~J., {Elmore}, C., {Kliem}, B., {T{\"o}r{\"o}k}, T., \& {Title},
  A.~M. 2008, \apj, 674, 586

\bibitem[{{Schrijver} \& {Title}(2011)}]{schrijver11}
{Schrijver}, C.~J., \& {Title}, A.~M. 2011, Journal of Geophysical Research
  (Space Physics), 116, A04108

\bibitem[{{Syrovatskii}(1982)}]{syrovatskii82}
{Syrovatskii}, S.~I. 1982, \solphys, 76, 3

\bibitem[{{Titov} \& {D{\'e}moulin}(1999)}]{titov99}
{Titov}, V.~S., \& {D{\'e}moulin}, P. 1999, \aap, 351, 707

\bibitem[{{Titov} {et~al.}(2002){Titov}, {Hornig}, \& {D{\'e}moulin}}]{titov02}
{Titov}, V.~S., {Hornig}, G., \& {D{\'e}moulin}, P. 2002, Journal of
  Geophysical Research (Space Physics), 107, 1164

\bibitem[{{Titov} {et~al.}(2011){Titov}, {Miki{\'c}}, {Linker}, {Lionello}, \&
  {Antiochos}}]{titov11}
{Titov}, V.~S., {Miki{\'c}}, Z., {Linker}, J.~A., {Lionello}, R., \&
  {Antiochos}, S.~K. 2011, \apj, 731, 111

\bibitem[{{T{\"o}r{\"o}k} {et~al.}(2009){T{\"o}r{\"o}k}, {Aulanier},
  {Schmieder}, {Reeves}, \& {Golub}}]{torok09}
{T{\"o}r{\"o}k}, T., {Aulanier}, G., {Schmieder}, B., {Reeves}, K.~K., \&
  {Golub}, L. 2009, \apj, 704, 485

\bibitem[{{T{\"o}r{\"o}k} {et~al.}(2010){T{\"o}r{\"o}k}, {Berger}, \&
  {Kliem}}]{torok10}
{T{\"o}r{\"o}k}, T., {Berger}, M.~A., \& {Kliem}, B. 2010, \aap, 516, A49

\bibitem[{{T{\"o}r{\"o}k} {et~al.}(2011){T{\"o}r{\"o}k}, {Chandra}, {Pariat},
  {D{\'e}moulin}, {Schmieder}, {Aulanier}, {Linton}, \& {Mandrini}}]{torok11}
{T{\"o}r{\"o}k}, T., {Chandra}, R., {Pariat}, E., {D{\'e}moulin}, P.,
  {Schmieder}, B., {Aulanier}, G., {Linton}, M.~G., \& {Mandrini}, C.~H. 2011,
  \apj, 728, 65

\bibitem[{{T{\"o}r{\"o}k} \& {Kliem}(2003)}]{torok03}
{T{\"o}r{\"o}k}, T., \& {Kliem}, B. 2003, \aap, 406, 1043

\bibitem[{{T{\"o}r{\"o}k} \& {Kliem}(2007)}]{torok07}
---. 2007, \an, 328, 743

\bibitem[{{T{\"o}r{\"o}k} {et~al.}(2004){T{\"o}r{\"o}k}, {Kliem}, \&
  {Titov}}]{torok04}
{T{\"o}r{\"o}k}, T., {Kliem}, B., \& {Titov}, V.~S. 2004, \aap, 413, L27

\bibitem[{{Vr{\v s}nak}(2008)}]{vrsnak08}
{Vr{\v s}nak}, B. 2008, Annales Geophysicae, 26, 3089

\bibitem[{{Wang} {et~al.}(2001){Wang}, {Chae}, {Yurchyshyn}, {Yang},
  {Steinegger}, \& {Goode}}]{wang01}
{Wang}, H., {Chae}, J., {Yurchyshyn}, V., {Yang}, G., {Steinegger}, M., \&
  {Goode}, P. 2001, \apj, 559, 1171

\bibitem[{{Wang} {et~al.}(2007){Wang}, {Sheeley}, \& {Rich}}]{wang07a}
{Wang}, Y., {Sheeley}, Jr., N.~R., \& {Rich}, N.~B. 2007, \apj, 658, 1340

\bibitem[{{Wheatland} \& {Craig}(2006)}]{wheatland06}
{Wheatland}, M.~S., \& {Craig}, I.~J.~D. 2006, \solphys, 238, 73

\bibitem[{{Zhukov} \& {Veselovsky}(2007)}]{zhukov07}
{Zhukov}, A.~N., \& {Veselovsky}, I.~S. 2007, \apjl, 664, L131

\bibitem[{{Zuccarello} {et~al.}(2009){Zuccarello}, {Romano}, {Farnik},
  {Karlicky}, {Contarino}, {Battiato}, {Guglielmino}, {Comparato}, \&
  {Ugarte-Urra}}]{zuccarello09}
{Zuccarello}, F., {Romano}, P., {Farnik}, F., {Karlicky}, M., {Contarino}, L.,
  {Battiato}, V., {Guglielmino}, S.~L., {Comparato}, M., \& {Ugarte-Urra}, I.
  2009, \aap, 493, 629

\end{thebibliography}
\end{document}